\documentclass[aps,pra,twocolumn,showpacs]{revtex4}
\usepackage{graphicx}
\usepackage{bm}
\usepackage{amsfonts}
\usepackage{amssymb}
\usepackage{amsmath}
\usepackage{times}
\usepackage{latexsym}
\usepackage{amsbsy}
\usepackage{pifont}
\usepackage{computer2}

\begin{document}

\title{Bound Modes in Dielectric Microcavities}
\author{P.M.\ Visser, K.\ Allaart, and D.\ Lenstra}
\affiliation{Vrije Universiteit Amsterdam, De Boelelaan 1081, 1081HV Amsterdam, The Netherlands}
\email{PMV@nat.VU.nl, Allaart@nat.VU.nl, Lenstra@nat.VU.nl}

\begin{abstract}
We demonstrate how exactly bound cavity modes can be realized in dielectric structures other than $3$d photonic crystals. For a microcavity consisting of crossed anisotropic layers, we derive the cavity resonance frequencies, and spontaneous emission rates. For a dielectric structure with dissipative loss and central layer with gain, the $\beta$ factor of direct spontaneous emission into a cavity mode and the laser threshold is calculated.
\end{abstract}

\maketitle

\section{Introduction}

One of the motivations for making a small and efficient laser is to enter a regime where quantum correlations in the emitted light can be created and studied. For this purpose spontaneous photons must be emitted into the cavity mode with high efficiency (the $\beta$ factor), so that the noise from the random emission in other modes is small \cite{Meystre,Yamamoto}. At the same time, the lifetime of the lasing mode needs to be large enough, to allow for stimulated emission. The fabrication of dielectric microstructures, like microspheres \cite{Eschmann,Sandoghdar} and VCSELs (vertical cavity surface-emitting lasers) \cite{Gerard,DeppeGraham,Forchel,Yamamoto2} for the realization of strong coupling and high finesse is therefore a big challenge for cavity QED-type experiments. From a fundamental point of view this also raises the question: what are the different types of systems where a small mode volume is still compatible with little leakage? Neither an ideal sphere \cite{Ching,Chang} nor a stack of disks (an ideal VCSEL), for example, give rise to mathematically bound solutions. These systems have resonances with a finite width, which only vanishes for large dimensions. When these structures have dimensions of the order of one optical wavelength, the field strength of the mode may become large (and thereby the coupling constant), but at the cost of an increasing loss rate. Systems with exact bound states can exist in so-called photonic materials. The well known examples are photonic crystals with a point defect \cite{Yablonovitch,Joannopoulos} and disordered structures \cite{Anderson,John,Lagendijk}. These systems are interesting for cavity QED, because ideally (no dissipation and infinite size) the lifetime of the bound state is infinite and propagating modes occur at other frequencies only. Photonic materials are complicated structures, however, so that fabrication and modeling is not easy. In this paper we demonstrate the occurrence of exact $3$d-bound states in dielectric structures that are not a photonic crystal.

\section{Bound Modes in Anisotropic Dielectric}

The system with the simplest design is shown in Figure \ref{Fig1}. This cavity consists of horizontal layers reminiscent of a VCSEL, with two crossed vertical planes of a material with lower index of refraction. The occurrence of bound states relies on the use of dielectric material with particular polarization properties. As we will show, the structure of Fig.\ \ref{Fig1} is in fact only one in a class of systems with bound modes. The systems in this class are characterized by the dielectric function $\varepsilon(\vec{r})$ with the tensor form
\begin{equation}
\varepsilon(\vec{r})/\varepsilon_1 = 1+\hat{z}\,\hat{z}[U(x) + V(y)] + 
(1-\hat{z}\,\hat{z}) W(z) .
\label{tenschi}
\end{equation}
Here $\varepsilon_1$ is the dielectric constant of the (isotropic) background medium. The functions $U$, $V$, and $W$ represent the spatial dependence of the vertical and horizontal layered structures. In the case that the layers have a lower susceptibility than the background (for instance, when they represent air holes in dielectric material), $U$, $V$ or $W$ will be negative. These functions will be specified later. In a dielectric environment of the form (\ref{tenschi}), fields with a polarization in the horizontal plane will only be affected by the structures described by $W(z)$, while vertically polarized fields are only perturbed by the structures described by $U(x)+V(y)$. Moreover, because the structures depend on one or two spatial coordinates only, the propagation of these particular fields in the other directions is free. Here we consider another specific type of polarization, which is governed by all three functions $U$, $V$, $W$. The electric field strength of this type can be expressed as
\begin{equation}
\vec{E}(\vec{r}) = \big( k_z^2\vec{\nabla}/k^2 - 
\hat{z}d/dz \big) f(x)g(y)h(z) ,
\label{vecE}
\end{equation}
in terms of scalar functions $f(x)$, $g(y)$, $h(z)$ and eigenvalues $k_z$, $k$. This is proven by direct substitution into the Maxwell equation $\vec{\nabla}\times\vec{\nabla}\times\vec{E} = (\omega/c)^2\varepsilon(\vec{r})\vec{E}$. After projecting out the Cartesian vector components, one finds that this equation separates into independent equations for the scalar functions:
\begin{eqnarray}
-(d/dx)^2 f(x) &=& k_x^2 f(x) + (k^2-k_z^2) U(x) f(x) ,
\label{Fx} \\
-(d/dy)^2 g(y) &=& k_y^2 g(y) + (k^2-k_z^2) V(y) g(y) ,
\label{Fy} \\
-(d/dz)^2 h(z) &=& k_z^2 h(z) + k_z^2 W(z) h(z) .
\label{Ez}
\end{eqnarray}
The separation constants $k_x$, $k_y$, and $k_z$ in these equations, which denote wave vector components, must be related by $k_x^2+k_y^2+k_z^2=k^2$, with the length of the wave vector given by $k=\varepsilon_1^{1/2}\omega/c$. It can now be verified directly, that the field (\ref{vecE}) is transversely polarized in the isotropic background. Note also that components normal to interfaces (where $U$, $V$ or $W$ have steps) of the electric field $\vec{E}$, the displacement field $\vec{D}=\varepsilon(\vec{r})\vec{E}$ and of the magnetic field $\vec{B}=-(i/\omega)\vec{\nabla}\times\vec{E}$ are all continuous.

That a dielectric structure of form (\ref{tenschi}) with specific choice of $U$, $V$, and $W$ can indeed support bound modes now follows from the following general considerations. Solutions that are localized in the $x$, $y$ or $z$ direction can be created when $k_x^2$, $k_y^2$ or $k_z^2$ is negative. Such solutions represent electromagnetic fields that are guided by vertical structures or a horizontal structure. Because Eqs.\ (\ref{Fx}-\ref{Fy}) are Schr\"odinger-type equations, it follows from standard wave mechanics that when $U$ and $V$ are positive in a finite region, the prefactor $k^2-k_z^2=k_x^2+k_y^2$ must be positive for such a guided wave. In that case both functions $U(x)$ and $V(y)$ act as attracting potentials, but only one of the two eigenvalues $k_x^2$, $k_y^2$ is negative. When, on the other hand, $U$ and $V$ both become negative, an attractive potential is obtained for negative $k^2-k_z^2$, so that the sign in front of both functions is reversed. Therefore a wave can be localized both in the $x$ and $y$ directions simultaneously when $U$ and $V$ become less than $-1/2$. Eq.\ (\ref{Ez}), is of the Helmholz type, with the effective potential $W(z)$ multiplied by the eigenvalue. Only when $W$ becomes less than $-1$, an attractive potential with a localized solution occurs. Localized solutions for each of the scalar functions $f$, $g$ or $h$ can also be obtained when the corresponding function $U$, $V$ or $W$ is periodic in two half spaces. In that situation the wave function can be a defect state in a band gap. A fully bound state for the electric field (\ref{vecE}) will occur when all three scalar wave functions are localized simultaneously. This can be achieved by means of combination of the different localizing effects.

The conclusion of this analysis may be summarized as follows. When the structures described by $U$ and $V$ have a higher index of refraction than the background, the functions $U$ and $V$ are positive and a $3$d bound state requires periodic structures in one of the two horizontal directions and also in the vertical direction, i.e.\ the system must be a $2$d photonic crystal. When both $U$ and $V$ become less than $-1/2$, they describe crossed vertical slabs with a low index, and a $3$d bound state is found with a $1$d periodic structure of horizontal layers. When $W$ describes a horizontal layer of negative-index material, a $1$d photonic crystal of vertical layers is needed for a $3$d bound state. Because $k_x^2+k_y^2+k_z^2=k^2$, at least one of the eigenvalues $k_x^2$, $k_y^2$ or $k_z^2$ must be positive, so that localized solutions are not possible without the use of periodicity in at least one dimension. Because the type of systems proposed here is not a $3$d photonic crystal, the frequency of a bound state will ly in the continuum of propagating modes.

\section{Example: a 1d periodic structure}
\subsection{Mode Solutions}

\begin{figure}[lt]
\centerline{\includegraphics[width=7cm]{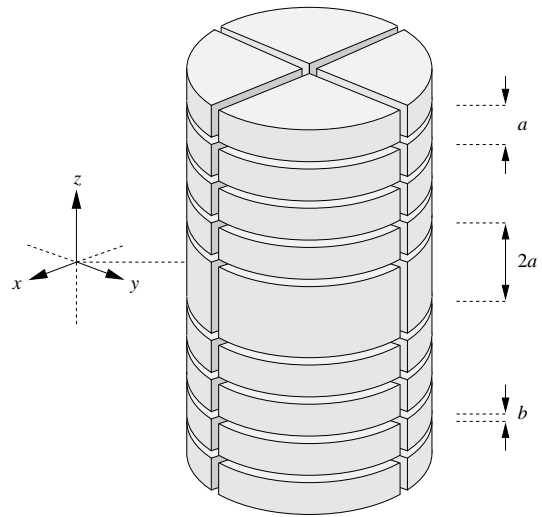}}
\caption[]{Example of an empty-cavity geometry where bound field modes are realized. For the vertical and horizontal layers, an anisotropic dielectric structure of the form of Eq.\ (\ref{tenschi}) is required, as discussed in the text below. The layers are depicted transparent, because their refractive index is lower than the index of the background. These layers, with their specific anisotropy, could be fabricated of air holes. The cylindrical boundary is not essential, because the bound modes decay exponentially with the distance from the origin.}
\label{Fig1}
\end{figure}

We will now consider the most interesting situation: the $1$d periodic structure illustrated in Fig.\ \ref{Fig1}, which has a $3$d bound state. The vertical and horizontal layers have widths $d$ and $b$ respectively and the layer-to-layer separation distance is $a$, so that the fundamental resonant wavelength $2\pi/k_z$ will lie in the interval between $2a$ and $4a$. The anisotropy inside the vertical layers is described with the dielectric tensor $\varepsilon=(1-\hat{z}\,\hat{z})\varepsilon_1 + \hat{z}\,\hat{z}\varepsilon_2$, while the tensor $\varepsilon=\hat{z}\,\hat{z}\varepsilon_1+(1-\hat{z}\,\hat{z})\varepsilon_2$ describes the horizontal layers. The (real part of the) dielectric constant $\varepsilon_2$ must be smaller than $\varepsilon_1$. This kind of anisotropy can be obtained by drilling air holes in isotropic high-index material in, respectively, the vertical and horizontal directions. We will discuss the cavity resonances of this structure and include absorption in the layers by allowing an imaginary component of the dielectric function. The exact bound states of our system must then arize from the resonances in the limit of zero loss. This idea is illustrated in the thin-layer approximation, neglecting the overlap regions (intersections of two crossing layers) in order to obtain analytical expressions.

The functions $U$ and $V$ in Eq.\ (\ref{tenschi}) describe the vertical low-index layers of width $d$ and are expressed in terms of the parameter $\chi=(\varepsilon_2-\varepsilon_1)d/\varepsilon_1$. The horizontal layers with thickness $b$ are described with $W$ and are likewise expressed in the parameter $\xi=(\varepsilon_2-\varepsilon_1)b/\varepsilon_1$. We adopt the forms
\begin{eqnarray}
U(x) &=& \theta(d-2|x|)\chi/d , \;\; V(y) = \theta(d-2|y|)\chi/d ,
\label{Vx} \\
W(z) &=& \sum_{l=-\infty}^\infty \theta(b-2|z-la|)\xi/b + \theta(b-2|z|)(\alpha-\xi)/b .
\label{Wz}
\end{eqnarray}
where $\theta(x)$ the Heaviside function. The central horizontal layer at $z=0$ has a different parameter $\alpha=(\varepsilon_3-\varepsilon_1)b/\varepsilon_1$. The situation $\alpha=0$ describes the empty cavity of Fig.\ \ref{Fig1}. We consider ${\rm Im}\,\chi>0$, ${\rm Im}\,\xi>0$ to model dissipative structures and ${\rm Im}\,\alpha<0$ to represent a central layer with gain. It follows from continuity and the Bloch theorem that the desired solutions of Eqs.\ (\ref{Fx}-\ref{Ez}) with potentials given by (\ref{Vx}-\ref{Wz}) are, outside the thin layers, of the form
\begin{eqnarray}
f(x) &=& e^{ik_x|x|}, \;\; g(y) = e^{ik_y|y|} ,
\nonumber \\
h(z) &=& e^{i(l+1)pa}\sin k_z(|z|-la) + e^{ilpa}\sin k_z(la+a-|z|) ,
\label{h(z)}
\end{eqnarray}
with $l={\rm int}(|z|/a)$ the number of layers between position $z$ and the origin. (Note that $|z|-la$ and $la+a-|z|$ are the distances to the nearest layer below and above $z$.) The solution $h(z)$ is an even function of $z$ and for $z>b/2$ one has $h(z+a)=e^{ipa}h(z)$, so that $p$ acts as a quasi momentum. Because the system is open and $\chi$, $\xi$, $\alpha$ are not real, $k_x$, $k_y$, $k_z$, $k$, and $p$ will generally be complex valued. The condition for an outgoing resonance implies ${\rm Re}\,k_x>0$, ${\rm Re}\,k_y>0$, and ${\rm Re}\,p>0$. These wave vector components can be expressed in terms of $k_z$ and the material constants by
\begin{eqnarray}
&& k_x = k_y = 2i/(-2\chi-d) ,
\nonumber \\
&& k^2 = k_z^2 - 8/(2\chi+d)^2 .
\label{k} \\
&& e^{ipa} = \cos k_z a - (k_z\alpha/2)\sin k_z a ,
\nonumber
\end{eqnarray}
The values of $k_z$ are the solutions of the closed equation
\begin{equation}
(2/k_z)^2[1 + \xi k_z\cot k_z a] = \xi^2 - (\xi-\alpha)^2 .
\label{sqrt}
\end{equation}
Modes that decay in time (${\rm Im}\,k<0$), describe loss of energy both by leakage of light out of the system to infinity and by absorption of the light in the structure. The amplified modes (${\rm Im}\,k>0$) are not realistic for long times, because saturation effects can not be neglected at high intensities. The spatially localized solutions describe the situation that the energy created at $z=0$ is fully dissipated in the other layers. In the limit of a passive cavity without loss, one has $\alpha={\rm Im}\,\chi={\rm Im}\,\xi=0$, so that the right-hand side of Eq.\ (\ref{sqrt}) disappears. Then the $k_z$ are real and ly inside the energy gaps of the band structure for the fully periodic potential $W(z)$ (the case $\alpha=\xi$). Because $k_x$ and $k_y$ are purely imaginary and $k_z$ is real, $k$ is real. This identifies the $3$d-bound states. Outside the layers the field is exponentially decaying in all directions. The decay in the $z$ direction in Eq.\ (\ref{h(z)}) derives from the quasi momentum of the form $p_n=n\pi/a+iq_n$ (the index $n=1,2,\ldots$ labels the resonances). The solutions to lowest order in $\xi$ are given by $k_{zn}=n\pi(a-\xi)/a^2$ and $q_n=(n\pi\xi)^2/2a^3$.

We give two numerical examples for the geometry $d=a=4b$. For $\varepsilon_1=13$, $\varepsilon_2=1$ (air holes in GaAs), one finds a period of $a=0.46\lambda$, ($\lambda$ is the wavelength of the first resonance) and $q_1a=0.29$. When $\varepsilon_1=2.3$, $\varepsilon_2=-20$ (silver in glass), then $a=0.68\lambda$, $q_1a=0.80$.

\subsection{Spontaneous emission factor and laser threshold}

\begin{figure}[lt]
\centerline{\includegraphics[width=8cm]{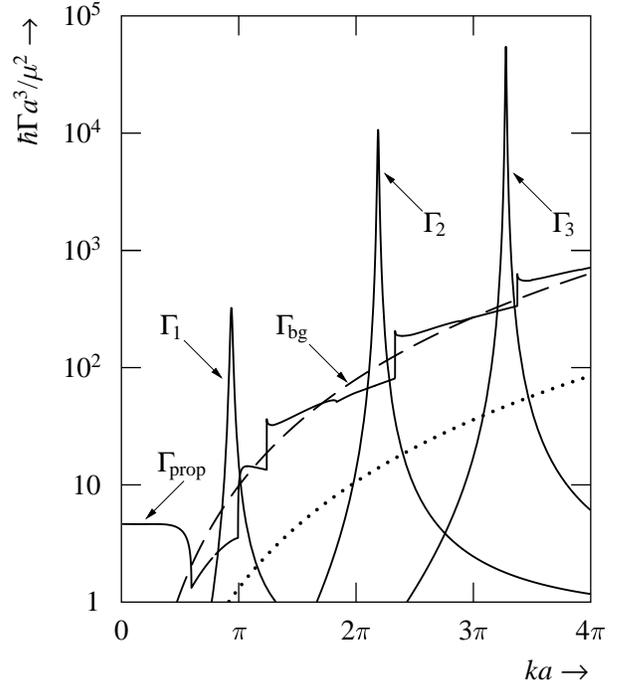}}
\caption[]{Spontaneous emission rates $\Gamma\!_n$ and $\Gamma\!_{\rm prop}$ into a cavity resonance and into the propagating modes, for a horizontal dipole emitter in the center of the geometry of Fig.\ \ref{Fig1}, as a function of the wave vector (the frequency is $\omega=ck/\varepsilon_1^{1/2}$). The adopted effective $2$d dielectric susceptibilities are $\chi=-12a/13$, $\xi=-3a/13$, with absorptive parts $\eta=0.01\chi$, $\zeta=0.01\xi$. This could correspond to $d=a=4b$ and a background dielectric constant $\varepsilon_1=13$ and with $\varepsilon_2=1$. The dashed and dotted curves are the emission rates in the back ground and in vacuum. The width of the sharp resonances ($n=1,2,3$) vanishes for $\zeta=0$ and when the absorptive loss is compensated by gain in an active layer at $z=0$.}
\label{Fig2}
\end{figure}

\begin{figure}[lt]
\centerline{\includegraphics[width=8cm]{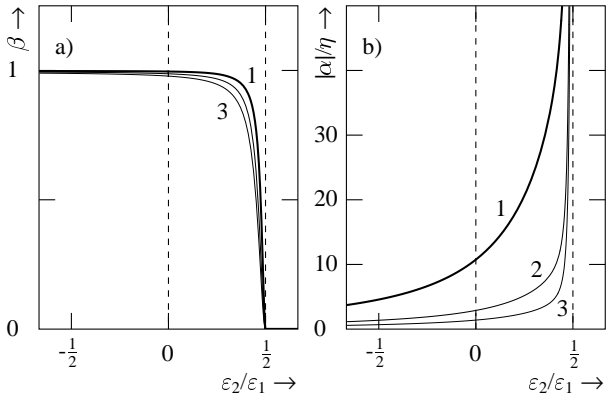}}
\caption[]{a) Spontaneous emission factor $\beta$ into the cavity mode versus the dielectric constant of the structure $\varepsilon_2$, at the first three resonant values $k_n$ ($n=1,2,3$). These correspond to the maxima of the peaks in Fig.\ \ref{Fig2}. b) Gain constant $\alpha$ versus dielectric constant $\varepsilon_2$ that is required to reach the lasing threshold, according to Eq.\ (\ref{alpha}). The absorptive components of the effective dielectric susceptibility were taken as $\eta=0.01\chi$, $\zeta=0.01\xi$.}
\label{Fig3}
\end{figure}

In presence of absorption, the imaginary part of the frequency $\omega_n=ck_n/\varepsilon_1^{1/2}$ of an outgoing resonance determines the decay rate of the probability that a photon is present in the corresponding mode. For small values of the absorption in the layers and for small gain, the following approximation applies. Define from here on the real and (small) imaginary parts of the effective $2$d susceptibilities and the wave vector with the notation $\chi+i\eta$, $\xi+i\zeta$ and $k_n-i\gamma_n$. The small decay rate of a resonance is then $c\gamma_n/\varepsilon_1^{1/2}$. This can be interpreted as an imaginary correction term to the real frequency $\omega_n$ of the $3$d-bound state. The result of the expansion of Eqs.\ (\ref{k}-\ref{sqrt}) is
\begin{equation}
\gamma_n = \frac{(k_{zn}^2-k_n^2)^{3/2}\eta}{k_n\sqrt{2}} + 
\frac{k_{zn}^2\zeta-ik_{zn}^4\xi^2\alpha/2}{(\xi+a+k_{zn}^2\xi^2 a)k_n} .
\label{gamma}
\end{equation}
Before including a linear gain medium at $z=0$, we first consider spontaneous emission rates from a single dipole placed at the origin of the cavity. The probability that a spontaneous photon is emitted directly into a cavity resonance is $\beta_n=\Gamma\!_n/(\Gamma\!_n+\Gamma\!_{\rm prop})$ in terms of the partial rate into a cavity resonance $\Gamma\!_n$ and into the total of propagating modes $\Gamma\!_{\rm prop}$. The contribution of the resonances to the emission rate is given by a sum of the Lorentzians
\[
\Gamma\!_n(k) = (6\pi\Gamma/R_n)\gamma_n \big/ [(k-k_n)^2+\gamma_n^2] .
\]
Here, $R_n=(1/\xi\!+\!k_{zn}^2 a/\sin\!^2 k_{zn} a)(4\chi^2\!-\!d^2)(k_n/k_{zn})^4$ is the intensity of the (normalized) bound state at the origin, $\Gamma=12k^3\mu^2\varepsilon_1^{5/2}/\hbar(2\!+\!\varepsilon_1)^2$ is the emission rate in the background medium \cite{Glauber}, and the width $\gamma_n$ is given by Eq.\ (\ref{gamma}) for $\alpha=0$. The rates are plotted in Fig.\ \ref{Fig2}, as a function of the wave vector for a fixed value of the effective susceptibilities $\chi$, $\xi$. The efficiency factor at the resonances is plotted in Fig.\ \ref{Fig3}a). There are several types of propagating modes; their classification and specific features will be discussed elsewhere \cite{Visser&Lenstra}. For strong coupling, $c\gamma_n/\varepsilon_1^{1/2}\ll\Gamma$, the decay is no longer exponential \cite{CohenTannoudji} and one enters the cavity QED regime, \cite{Jaynes,Zhu,Bernardot,Thompson,Norris}.

For the parameter ${\rm Im}\,\alpha>0$ our model represents a cavity with a gain layer at $z=0$. This gain compensates the loss in the layers and reduces the width of the cavity resonances, as can be seen from Eq.\ (\ref{gamma}). When the loss is fully compensated, the stationary situation $\gamma_n=0$ arizes. This corresponds to the lasing threshold. The gain $\alpha$ needed to compensate for a small loss, is linear in the loss constants $\eta$, $\zeta$. From Eq.\ (\ref{gamma}) it follows that the required value of $\alpha$ is given, including the effect of the beta factor, by
\begin{equation}
\alpha = \frac{-2i}{\beta_n}\, \Big[ \frac{\zeta}{k_{zn}^2\xi^2} + \frac{(k_{zn}^2-k_n^2)^{3/2}\eta}{\sqrt{2}}\, \frac{\xi+a+k_{zn}^2\xi^2 a}{k_{zn}^4\xi^2} \Big] .
\label{alpha}
\end{equation}
The behavior as a function of the susceptibility $\xi$ is plotted in Fig.\ \ref{Fig3}b). We conclude that the required value of the effective gain constant $\alpha$ can easily be reached for semiconductor material (for which the value $\alpha=-10i\eta$ is quite realistic). The behavior of the laser above threshold is not discussed here, because then the stationary state is determined by saturation, not included in our model.

\section{Conclusion}

Apart from absorption, a structure of finite size (layers of finite extent and finite in number) will also give rise to leakage. This, of course, is also the case for a defect state in a common three-dimensional photonic crystal. The finite size approximation is likely to be very good, however, since the (analytic) mode functions are exponentially localized evanescent waves. Therefore, the coupling to the outside world is exponentially small in the system size, while the mode volume remains of the order of one cubic wavelength. For a cubic system of size $Nd\times Nd\times Na$, the fraction of mode volume outside the system is roughly $\exp(-2N|k_x|d-Nq_1a)$. For $N>6$ (GaAs case), and for $N>23$ (glass/silver case), this fraction is less than $10^{-10}$. In this respect, the proposed device of Figure \ref{Fig1} provides an interesting alternative to the use of whispering gallery modes in dielectric spheres, where smaller mode volume leads to increasing loss. Semiconductor structures with $3$d-bound states may be very promising for future cavity QED experiments and quantum communication applications.

\acknowledgements

This work is part of the research program of the `Stichting voor Fundamenteel Onderzoek der Materie' (FOM), which is financially supported by the `Nederlandse Organisatie voor Wetenschappelijk Onderzoek' (NWO).


\begin{references}
\small
\bibitem{Meystre}
R.\ Jin, D.\ Boggavarapu, M.\ Sargent, P.\ Meystre, H.M.\ Gibbs, G.\ Khitrova, 
\pra {\bf 49}, 4038 (1994).
\bibitem{Yamamoto}
Y.\ Yamamoto, S.\ Machida, G.\ Bj\"ork, \pra {\bf 44}, 657 (1991);
Y.\ Yamamoto, S.\ Machida, Y.\ Horikoshi, K.\ Igeta, G.\ Bj\"ork, \oc {\bf 80}, 
337 (1991);
G.\ Bj\"ork, H.\ Heitmann, Y.\ Yamamoto, \pra {\bf 47}, 4451 (1993).
\bibitem{Eschmann} 
A.\ Eschmann, C.W.\ Gardiner, \pra {\bf 49}, 2907 (1994).
\bibitem{Sandoghdar}
V.\ Sandoghdar, F.\ Treussart, J.\ Hare, V.\ Lef\`evre-Seguin, J.-M.\ Raimond, 
S.\ Haroche, \pra {\bf 54}, R1777 (1996).
\bibitem{Gerard} 
J.M.\ G\'erard, B.\ Sermage, B.\ Gayral, B.\ Legrand, E.\ Costard, V.\ Thierry-Mieg, \prl {\bf 81}, 1110 (1998).
\bibitem{DeppeGraham} 
D.\ Deppe, L.\ Graham, D.\ Huffaker, J.\ of Quant.\ Electron.\ {\bf 35}, 1502 (1999).
\bibitem{Forchel}
M.\ Bayer, T.L.\ Reinecke, F.\ Weidner, A.\ Larionov, A.\ McDonald, A.\ Forchel, \prl {\bf 86}, 3168 (2001).
\bibitem{Yamamoto2}
G.S.\ Solomon, M.\ Pelton, Y.\ Yamamoto, \prl {\bf 86}, 3903 (2001).
\bibitem{Ching}
S.C.\ Ching, H.M.\ Lai, K.\ Young, \josa {\bf B 4}, 1995 (1987);
S.C.\ Ching, H.M.\ Lai, K.\ Young, \josa {\bf B 4}, 2004 (1987);
\bibitem{Chang}
P.T.\ Leung, S.Y.\ Liu, K.\ Young, \pra {\bf 49}, 3057 (1994);
E.S.C.\ Ching, P.T.\ Leung, K.\ Young in {\it Optical Processes in 
Microcavities}, editors: R.K.\ Chang, A.J.\ Campillo, (World Scientific 1996), 
p.\ 1.
\bibitem{Yablonovitch} 
E.\ Yablonovitch, \prl {\bf 58}, 2059 (1987).
\bibitem{Joannopoulos}
P.R.\ Villeneuve, S.\ Fan, J.D.\ Joannopoulos, \prb {\bf 54}, 7837 (1996).
\bibitem{Anderson}
P.W.\ Anderson, Phys.\ Rev.\ {\bf 109}, 1493 (1958);
P.W.\ Anderson, Philos.\ Mag.\ B {\bf 52}, 505 (1985).
\bibitem{John}
S.\ John, \prl {\bf 53}, 2169 (1984).
\bibitem{Lagendijk}
A.\ Lagendijk, M.P.\ van Albada, M.B.\ van der Mark, Physica (Amsterdam) {\bf 
104A}, 183 (1986).
\bibitem{Glauber}
R.J.\ Glauber, M.\ Lewenstein, \pra {\bf 43}, 467-491 (1991).
\bibitem{Visser&Lenstra}
P.M.\ Visser, K.\ Allaart, D.\ Lenstra, to appear in \pre {\bf 65} (2002).
\bibitem{CohenTannoudji}
C.\ Cohen-Tannoudji, J.\ Dupont-Roc, G.\ Grynberg, {\it Atom-Photon 
Interactions}, (John Wiley \& Sons, New York 1992).
\bibitem{Jaynes}
E.T.\ Jaynes, F.W.\ Cummings, Proc.\ IEEE {\bf 51}, 89 (1963).
\bibitem{Zhu}
Y.\ Zhu, D.J.\ Gaultier, S.E.\ Morin, Q.\ Wu, H.J.\ Carmichael, T.W.\ Mossberg, 
\prl{\bf 64}, 2499 (1990).
\bibitem{Bernardot}
F.\ Bernardot, P.\ Nussensveig, M.\ Brune, J.M.\ Raimond, S.\ Haroche, 
Europhys.\ Lett.\ {\bf 17}, 33 (1992).
\bibitem{Thompson}
R.J.\ Thompson, G.\ Rempe, H.J.\ Kimble, \prl {\bf 68}, 1132 (1992).
\bibitem{Norris}
T.B.\ Norris, J.-K.\ Rhee, C.-Y.\ Sung, \prb {\bf 50}, 14663 (1994).
\end{references}
\end{document}